# Superconductivity of interface layer at contact between normal metal and high temperature superconductor

Oleg P. Ledenyov and Valery A. Frolov

*National Scientific Centre Kharkov Institute of Physics and Technology, Academicheskaya 1, Kharkov 61108, Ukraine.*

In this research, it is shown that there are the necessary physical conditions to originate the returning superconductivity in the thin interface layer at the contact between the normal metal and the high temperature superconductor (*N-S* contact). The influences by the temperature *T*, magnetic field *H* and direct current *I* on the electrical resistance $R_G$ of the thin interface layer *G* in the multilayered system *N-G-S*, where *N* can be one the normal metal layers of Argentum (*Ag*), Indium (*In*), Gallium-Indium (*Ga-50%In*), *G* is the thin interface layer, *S* is the high temperature superconductor (*HTS*) layer of $YBa_2Cu_3O_{7-x}$, are researched experimentally.

## Introduction

It is a well known fact that the electron properties of surface layers of materials and the electron properties of insight layers of materials are different, depending on the physical features of energy spectrum of electron states. The interface region with the new electron properties appears in the case, when the two materials have a close contact with each other [1]. The interface layer *G* at the contact between the massive high temperature superconductor (*S*) and the deposited normal metal (*N*) belongs to the interface regions described above. The research on the electron properties of the interface layer *G* is in the scope of our research interest, because it can provide some useful insights on both the nature of high temperature superconductivity as well as the possibility of synthesis of new material structures with the special physical properties.

In the frames of our experimental research, the influences by the temperature *T*, magnetic field *H* and direct current *I* on the electrical resistance $R_G$ of the thin interface layer *G* in the system *N-G-S*, where *N* is the normal metal layer of Argentum (*Ag*), Indium (*In*), Gallium-Indium (*Ga-50%In*), *G* is the thin interface layer, *S* is the high temperature superconductor (*HTS*) layer of $YBa_2Cu_3O_{7-x}$, are researched.

## Synthesis of Samples for Experimental Measurements

The samples with the interface layer *G* were made by deposing the covering metallic thin layer of selected metal (*Ag, In, Ga-50%In*) on the massive high temperature superconductor substrate of $YBa_2Cu_3O_{7-x}$ with the application of different methods. The layer of *Ag* was deposited on the $YBa_2Cu_3O_{7-x}$ substrate, using the method of *Ag* paste burning at the temperature *T≈200°C* over the time period of *15 min*. The layer of *In* was deposited on the $YBa_2Cu_3O_{7-x}$ substrate, using the *In* liquid phase deposition method with the application of ultrasound at the temperature *T≈160°C*. The layer of *Ga-50%In* alloy was deposited on the $YBa_2Cu_3O_{7-x}$ substrate, using the *Ga-50%In* liquid phase deposition method with the application of mechanical friction by the sharp edge of lancet at the room temperature.

## Experimental Measurements Setup

The experimental measurements setup was developed to provide an opportunity to simultaneously measure the electric resistances of both the *HTS* substrate ($R_S$) and the interface layer ($R_G$) in the two cases, when *1)* the magnetic field *H* with the magnitude of up to *4 kOe* is applied, and *2)* without the magnet field *H* application. Making the qualitative estimation of the order of magnitude of measured electric resistivities, it is necessary to provide the information that the specific electric volume resistivity of superconducting sample is $\rho_S \approx 10^{-3}$ *Ohm·cm* in the case of $YBa_2Cu_3O_{7-x}$ at the temperature of *300 K*, the specific electric surface resistivity of interface layer is $\rho_G \approx 10^{-2}$ *Ohm·cm²* in the cases of all the metals (*Ag, In, Ga-50%In*) at the temperature of *300 K*.

## Experimental Measurements Results

In Fig. 1, the dependences of the relative resistance on the temperature $R_G/R_{G300}(T)$ for the interface layers *G* between the deposited normal metal of *Ag* and the high temperature superconductor substrate of $YBa_2Cu_3O_{7-x}$ for the two differently synthesized substrates. The first substrate of $YBa_2Cu_3O_{7-x}$ (1) had the dependence of the resistance on the temperature $R_S(T)$, which is typical for the metal; the second substrate of $YBa_2Cu_3O_{7-x}$ (2) had the dependence of the resistance on the temperature $R_S(T)$, which is typical for the semiconductor. In the both experimentally researched cases, the interface layers $G$ had the dependences of the resistances on the temperatures $R_G(T)$, which are typical for the semiconductor in the range of temperatures from $300\ K$ up to $92\ K$; then, the sharp decreases in the dependences $R_G(T)$ below the temperature of $92\ K$ were observed; and finally, at the further decrease of temperature, the shapes of the dependences $R_G(T)$ returned back to a type of the dependence $R_G(T)$, which is characteristic for the semiconductor. In the case, when the transition interface layer $G$ is created by the contact between the high temperature superconductor of $YBa_2Cu_3O_{7-x}$ and the metal of $In$ or the metallic alloy of $Ga-50\%In$, then all the dependences $R_G(T)$ have a type of the dependence $R_G(T)$, which is characteristic for the semiconductor, but with the much stronger increase of dependence $R_G/R_{G300}(T)$ in the range of low temperatures (It is assumed that the $HTS$ has the metallic dependence $R_S(T)$). It is necessary to note that there is a very weak anomaly of $R_G(T)$ below the critical temperature $T_C$ of substrate of $YBa_2Cu_3O_{7-x}$ in the case of $In$; and this anomaly was not detected in the case of $Ga-50\%In$.

In our opinion, all the variations of dependences of the electric resistance on the temperature $R_G(T)$ can be connected with the presence of the *Schottky barrier* on the normal metal – superconductor (*N-S*) boundary, which is evidently confirmed by the electron tunneling experiments [2]. Taking to the consideration the fact that the number of electrons per atom in the surface layer of deposited normal metals is quite different ($Ag$ has the one valence, $In$ and $Ga-50\%In$ have the three electrons per atom), then it is possible to suppose that the electrons, filling the hole states, make the changing influence on the electron properties of the superconductor's region, which is in close proximity to the deposited layer of normal metal (*Schottky barrier*). In the case of a big number of the electrons per atom in the deposited layer of normal metal, the filling of the hole states by the electron states in the conduction zone in the interface layer $G$ of the $HTS$ sample is more complete, and the electro-physical properties of the $HTS$ layer, which is in close proximity to the deposited layer of normal metal, are very different from the electro-physical properties of the massive $HTS$ sample. In particular, in the range of all the temperatures, the type of dependence of the electric resistance on the temperature $R_G(T)$ in the interface layer $G$ is similar to the type of dependence of the electric resistance on the temperature $R(T)$ in the semiconductors in Fig. 2.

In the case of the surface deposited metallic layer with the valence of one, the energy spectrum of the interface layer $G$ is less deformed, comparing to the above discussed case, as a result the appearing semiconducting properties are not so strong. Let us suppose that the anomalies in the dependence $R_G(T)$ below the critical temperature $T_c$ are connected with the superconductivity, which is characteristic for the interface layer's material, - this fact, generally speaking, still must be evidently confirmed - then, going from the presented point of view, the smaller deformation of energy spectrum will suppress the superconductivity phenomena to a smaller degree. This is observed in the experiment as shown in Figs. 1, 2. Of course, the sample no. 2, which has the very different physical properties of substrate, was not selected for the comparative analysis.

The following question arises: Why does the superconductivity become so weak at the decrease of the temperature $T$ in an analogy with the observations, which were made in the cases of both the magnetic superconductors [3] and the low temperature superconducting ceramics [4] before? The decrease of a number of charges in the conduction zone at the decrease of the temperature $T$, as shown on the dependence $R_G(T)$ (see Figs. 1, 2), may be one of possible reasons of an appearance of the phenomena, which is called the returning superconductivity. The similar theoretical mechanism of the $HTS$, connected with an origination of the returning superconductivity, is considered in [5].

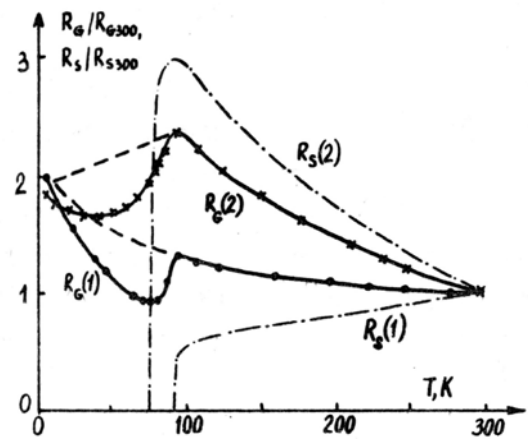

*Fig.1.* Dependences of electric resistances of two differently synthesized $YBa_2Cu_3O_{7-x}$ substrates no. 1 and no. 2 on temperature $R_s/R_{s300}$ (T) (See dashed lines with dots: $R_s(1)$ and $R_s(2)$).

Dependences of electric resistances on temperature $R_G/R_{G300}$ (T) of two interface layers $G$ in $YBa_2Cu_3O_{7-x}$ substrates no. 1 and no. 2 with deposited metallic layers of Ag (See solid lines: $R_G(1)$ and $R_G(2)$).

Dashed lines without dots denote for:
1) Dependence of normal electric resistance on temperature $R_G/R_{G300}$ (T) in sample no. 1, which is restored by current I ($YBa_2Cu_3O_{7-x}$ substrates no. 1 with deposited metallic layer of Ag);

2) Dependence of normal electric resistance on temperature $R_G/R_{G300}$ (T) in sample no. 2, which is restored by magnetic field H ($YBa_2Cu_3O_{7-x}$ substrates no. 2 with deposited metallic layer of Ag).

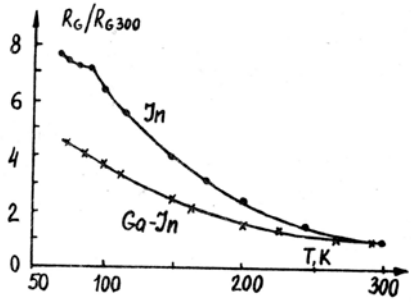

*Fig. 2. Dependences of electric resistances on temperature $R_G/R_{G300}$ (T) of two interface layers G in $YBa_2Cu_3O_{7-x}$ substrates no. 3 and no. 4 with deposited metallic layers of In and Ga-In50% correspondingly. Dependences of electric resistances on temperatures $R_s/R_{s300}$ (T) in $YBa_2Cu_3O_{7-x}$ substrates no. 3 and no. 4 correspond to curve $R_s(1)$ in Fig. 1.*

The finding of connection between the observed anomaly origination on the dependence $R_G(T)$ and the superconductivity phenomena is a scientific problem, which was solved by the completion of research toward the study of influence by the magnetic field $H$ on the electric resistance $R_G$ in the sample no. 2 at the different temperatures $T$. For this purpose, the interface layer $G$ was placed between the two poles of an electrical magnet, then the dependences $R_G(T)$ were measured at the constant temperature $T$ as shown in Fig. 3. It was found that the electric resistance $R_G$ of the substrate of $YBa_2Cu_3O_{7-x}$ at the temperatures $T$ below the critical temperature $T_c$ increases with an increase of magnitude of the magnetic field $H$, saturating at the magnetic field $H = 3$ kOe. This fact allows us to state that there is the superconductivity in the interface layer $G$. The experimental measurements, which were conducted on the resistive transitions in the interface layer $G$ in the sample no. 2 can be considered as another evidence that there is the superconductivity in the interface layer $G$ as shown in Fig. 4. The comparative analysis between the sample no. 1 and the sample no. 2, which have the deposited metallic layers of $Ag$ on the different substrates of $YBa_2Cu_3O_{7-x}$, then, it is possible to make a conclusion that they have the absolutely different electron properties at $T < 92$ K.

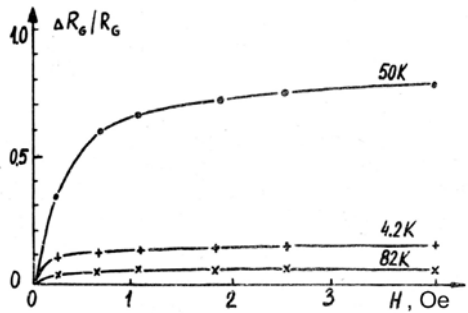

*Fig. 3. Dependence of relative increment of electric resistance $\Delta R_G/R_G = [R_G(T,H) - R_G(T,0)]/R_G(T,0)$ on magnetic field H in interface layer G in $YBa_2Cu_3O_{7-x}$ substrate no. 2 with deposited metallic layer of Ag. Dependence of electric resistance on temperature $R_s/R_{s300}$ (T) in $YBa_2Cu_3O_{7-x}$ substrate no. 2 correspond to curve $R_s(2)$ in Fig. 1.*

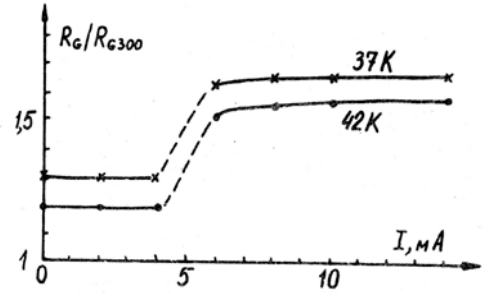

*Fig. 4. Dependence of relative electric resistance on magnitude of direct electric current $R_G/R_{G300}$ (I), which flows perpendicularly to interface layer G in $YBa_2Cu_3O_{7-x}$ substrate no. 1 with deposited metallic layer of Ag. Dependence of electric resistance on temperature $R_s/R_{s300}$ (T) in $YBa_2Cu_3O_{7-x}$ substrate no. 1 correspond to curve $R_s(1)$ in Fig. 1.*

Discussing the experimental results, let us explain that, in Fig. 1, the dashed lines denote for:
1) The dependence of the normal electric resistance on the temperature $R_G/R_{G300}$ (T) in, which is restored by the current $I$ ($YBa_2Cu_3O_{7-x}$ substrates no. 1 with deposited metallic layer of $Ag$);
2) The dependence of the normal electric resistance on the temperature $R_G/R_{G300}$ (T), which is restored by the magnetic field $H$ ($YBa_2Cu_3O_{7-x}$ substrates no. 2 with deposited metallic layer of $Ag$).

As far as the first curve is concerned, it can be seen that the semiconducting character of the dependence of the normal electric resistance on the temperature $R_G(T)$ in the sample no. 1 below the temperature $T$ of $92$ K is preserved. In the case of the interface layer $G$ in the sample no. 2 at the temperature $T$ of $92$ K, the quite unexpected fact is that the dependence of the normal electric resistance on the temperature $R_G(T)$ has the linear metallic character, allowing to make a conclusion that the semiconductor – metal electron transition in the system of charge carriers in the sample no. 2 at the temperature $T \approx 92$ K is present. Possibly, in the sample no. 2, the metallic character of the dependence of the normal electric resistance on the temperature $R_G(T)$ is observed, because the superconductivity in the interface layer $G$ is not suppressed up to the Helium temperatures. The real reasons of both *1)* the presence of the semiconductor – metal electron transition in the system of charge carriers in the sample no. 2 at the temperature $T \approx 92$ K, and *2)* the absence of the semiconductor – metal electron transition in the system of charge carriers in the sample no. 1 at the temperature $T \approx 92$ K, are not clear yet. In this connection, we would like to note that the observation of big variety of physical properties of interface layers can be somehow connected with the magnitude of their resistivity. As we have shown above, the contacts of $Ag$ - $YBa_2Cu_3O_{7-x}$ with the magnitude of resistivity $\rho \approx 10^{-2}$ $Ohm \cdot cm^2$ have the properties of the returning superconductivity. At the same time, in agreement with the literature data [6], the same contacts of $Ag$- $YBa_2Cu_3O_{7-x}$ with the more bigger magnitude of resistivity $\rho \approx 0,15$ $Ohm \cdot cm^2$ have the pure semiconducting dependence of the normal electric resistance on the temperature $R_G(T)$. Moreover, we report the strong influence by the previous sample treatment history on the physical behavior of the dependence of the normal electric resistance on the temperature $R_G(T)$. In particular, the fast thermo-cycling from the nitrogen temperature to the room temperature decreases the magnitude of the minimum in the dependence $R_G(T)$.

## Conclusion

The obtained experimental results show that there are all the necessary physical conditions to originate the returning superconductivity in the interface layer $G$ at the contact between the normal metal and the high temperature superconductor (*N-S* contact). We obtained the experimental dependences, which accurately characterize the influences by the temperature $T$, magnetic field $H$ and direct current $I$ on the electrical resistance $R_G$ of the thin interface layer $G$ in the system *N-G-S*, where *N* can be one of the normal metal layers of Argentum (*Ag*), Indium (*In*), Gallium-Indium (*Ga-50%In*); *G* is the thin interface layer; *S* is the high temperature superconductor (*HTS*) layer of $YBa_2Cu_3O_{7-x}$.

This experimental research is completed in the frames of the fundamental and applied superconductivity research program at the National Scientific Centre Kharkov Institute of Physics and Technology (*NSC KIPT*) in Kharkov in Ukraine.

Authors sincere thank Boris G. Lazarev for the thoughtful discussions on the obtained experimental research results and his enormous support of experimental research.

This research paper was published in the *Problems of Atomic Science and Technology* (*VANT*) in 1989 in [7].

*E-mail: ledenyov@kipt.kharkov.ua

———————